\documentclass[12pt,preprint]{aastex}

\shorttitle{Interstellar Iron and Silicon}
\shortauthors{Miller et al. 2007}

\begin{document}


\title{Interstellar Iron and Silicon Depletions in Translucent Sight Lines\footnote{Based on
observations with the NASA/ESA {\it Hubble Space Telescope},
obtained from the data archive at the Space Telescope Science Institute.
STScI is operated by the Association of Universities for Research in
Astronomy, Inc. under NASA contract NAS 5-26555.}}

\author{Adam Miller\footnote{Visiting student from the Massachusetts
Institute of Technology (MIT).}~$^,$\footnote{Current Address: Institute of
Astronomy, University of Cambridge, Madingley Road, Cambridge. CB3 0HA,
UK.}\ \ and J. T. Lauroesch\footnote{Current Address: Department of Physics and
Astronomy, University of Louisville, Louisville, KY 40292.}
}
\affil{Department of Physics and Astronomy, Northwestern University, 2145
Sheridan Road, Evanston, IL 60208}
\email{amiller@ast.cam.ac.uk, jtl@elvis.astro.northwestern.edu}

\author{Ulysses J. Sofia}
\affil{Department of Astronomy, Whitman College, 345 Boyer Avenue, Walla Walla, WA 99362}
\email{sofiauj@whitman.edu}

\author{Stefan I. B. Cartledge\footnote{Current Address: Department of Physics
        \& Astronomy, Valparaiso University, 1610 Campus Drive East,
        Valparaiso, IN 46383}}
\affil{Department of Physics and Astronomy, Louisiana State University, Baton Rouge, LA 70803}
\email{scartled@gmail.com}

\and

\author{David M. Meyer}
\affil{Department of Physics and Astronomy, Northwestern University, 2145
Sheridan Road, Evanston, IL 60208}
\email{davemeyer@northwestern.edu}


\begin{abstract}

We report interstellar \ion{Fe}{2} and \ion{Si}{2} column densities
toward six translucent sight lines ($A_{V} \ga 1$) observed with the
Space Telescope Imaging Spectrograph (STIS). The abundances were determined
from the absorption of \ion{Si}{2}] at 2335 \AA, and several weak Fe
transitions including the first reported detections of the 2234 \AA\
line.  We derive an empirical $f$-value for the \ion{Fe}{2} $\lambda 
2234$ \AA\ transition of log($f\lambda$) = $-1.54 \pm 0.05$.
The observed sight lines sample a variety of extinction characteristics as
indicated by their $R_{V}$ values, which range from 2.6 - 5.8. The dust-phase 
abundances of both Si and Fe are positively correlated with the
small-grain population (effective radii smaller than a few hundred
$\micron$) toward the targets. The physical conditions along the sight
lines suggest that this relationship may be due to differences in the
survival of small particles in some interstellar environments. The chemical
composition of the small grains could either resemble dust mantles or be silicate rich.

\end{abstract}


\section{Introduction}
Highly abundant species play a significant role in
the formation of interstellar dust grains, and thus are important to
our understanding of the chemical evolution of interstellar clouds. 
Interstellar silicon and iron abundances are two of the most important
diagnostics for studying dust. Only oxygen and carbon make up a higher
fraction of grain mass, and the large abundances and volatility of
these elements make it difficult to determine their precise dust-phase
abundances. Conversely, silicon and iron can provide us with detailed
information about the dust composition in sight lines that sample a wide
range of physical conditions. Silicates appear to be ubiquitous in the
neutral ISM \citep{vladilo2002,sofia2005b}, and iron always shows a
substantial depletion from the gas phase \citep{jenkins1987}. Thus silicon
and iron are fundamental grain constituents of most interstellar dust models
\citep{draine2003}.

The dominant ions of iron and silicon in the neutral ISM
are \ion{Fe}{2} and \ion{Si}{2}. 
Previous surveys of interstellar \ion{Fe}{2} and \ion{Si}{2} abundances
were performed with data from {\it Copernicus} \citep{jenkins1986} and the
{\it International Ultraviolet Explorer} ({\it IUE}) \citep{vansteenberg1988}
satellites.  These measurements suffered from the limitations (particularly 
background uncertainties and limited spectral resolution) of both instruments, and to a lesser degree from 
large uncertainties associated with the $f$-values for the
absorption lines used in these studies.  This situation was greatly
improved with the launch of the {\it Hubble Space Telescope} ({\it HST}), 
which has superior spectral resolution and sensitivity. Such data can be used to detect 
the extremely weak lines that are unlikely to suffer from saturation effects. 

In the past decade, new accurate laboratory $f$-values have been determined
for numerous transitions of important species including the \ion{Fe}{2} $\lambda\lambda$2249,
 2260 \AA\ lines \citep{bml1994} as well as the very weak inter-system
\ion{Si}{2}] $\lambda$2335 \AA\ line \citep{csb1993}. Despite these
improvements, very few reliable measurements of both the iron and
silicon abundance in
the same sight line have been made.  Additionally, as we begin to look at
higher column density sight lines even the (relatively) weak \ion{Fe}{2}
transitions at $\lambda\lambda$1142, 2249, 2260 \AA, which have well known
oscillator strengths, begin to suffer from saturation effects. Therefore,
even weaker lines such as the \ion{Fe}{2} transitions at $\lambda$2234 and
2367 \AA\ are important for an accurate determination of column densities.
Unfortunately the \ion{Fe}{2} transitions at $\lambda$2234 and 2367\AA\ do
not have experimentally derived $f$-values.

In this paper we use these very weak iron and silicon lines to investigate the 
translucent sight line abundances.
The goal is to better understand the dust characteristics of these sight
lines, which have previously been extensively studied for carbon, oxygen, and
krypton abundances \citep{cartledge2001, sofia2004}.
In $\S$2 we discuss the sample and the data. In $\S$3 we present the
iron and silicon depletions, and discuss the results in $\S$4.

\section{Observations and Data Reduction}

The six primary sight lines for this study were originally selected
for determining the
abundance of interstellar carbon \citep{sofia2004}.  A summary of the sight
line characteristics for our sample is given in Table 1. As discussed by
\citet{sofia2004}, the weakness of the interstellar 2325 \AA\ \ion{C}{2}]
feature required very high signal-to-noise ratios to be obtained with the
E230H echelle mode of STIS aboard the {\it Hubble Space Telescope}.  Within
the wavelength region covered by these observations are also located the
weak intersystem line of \ion{Si}{2}] at 2335 \AA\ and multiple
transitions of \ion{Fe}{2} including the very weak absorption lines at
2234 \AA\ and 2367 \AA. A detailed description of the data is given
in \citet{sofia2004}. 
As discussed in that paper, a number of tests were performed with these
datasets to obtain the most reliable measurements of weak lines. In addition to 
the {\it HST} spectra, {\it Far-Ultraviolet Spectroscopic Explorer (FUSE)}
data from \citet{cartledge2004} were used. Figure 1
shows the normalized Si ({\it solid line}) and Fe ({\it dotted line})
absorption features for each of the sight lines superimposed on different
vertical scales.

\section{Results}
\subsection{Ionization and Excitated States}

Based on the ionization potentials of neutral iron and silicon
(7.9 and 8.1 eV respectively), it is natural to assume that the dominant
state contributing to the iron and silicon columns in the \ion{H}{1} gas
in our six lines of sight would be the ground--state of \ion{Fe}{2} and
\ion{Si}{2}, which have ionization potentials of 16.2 and 16.3 eV respectively.
In deriving the gas--phase abundances of these elements one also
has to consider the low--lying excited--states of \ion{Si}{2} and \ion{Fe}{2}
which can be populated by collisions and/or photon pumping, in addition to
the neutral and doubly--ionized states.  We note that besides the
ground--state transitions of \ion{Si}{2} and \ion{Fe}{2},
transitions arising from \ion{Si}{1} and \ion{Fe}{1} and the
excited states of \ion{Si}{2} and \ion{Fe}{2}
are within our wavelength coverage or within the wavelength coverage of
the E140H ({\rm $\lambda_{cent}=1271$ \AA}) data obtained for these stars
discussed in \citet{cartledge2001}.   While the \ion{Si}{3} (1206\AA)
line is covered, its intrinsic strength makes it an unsuitable transition for
determination of the column density of doubly--ionized silicon.  Therefore
we will use measurements of the populations of neutral and singly--ionized
silicon and iron as well as evidence from other elements to determine if
the assumption that the ground state of singly--ionized iron
and silicon indeed dominate the columns in the \ion{H}{1} gas is correct.

We can derive the neutral columns from the \ion{Fe}{1} (2167,
2198\AA) and \ion{Si}{1} (2208\AA) lines in our datasets.  These
lines are generally very weak, indeed the largest equivalent width
for \ion{Fe}{1} (2167\AA) is detected toward HD~147888 at 15.7$\pm$0.9
m\AA, corresponding to a column density of ${\rm 3\times 10^{12} cm^{-2}}$,
or $\sim$0.1\% of the \ion{Fe}{2} column.  Even toward HD~27778, with the
lowest column of \ion{Fe}{2} in our sample, the neutral fraction is less
than 1\%.  For \ion{Si}{1} the line at 2208.67 \AA\ has an oscillator
strength of $5.75\times 10^{-2}$ (Morton 2003), a factor of
$1.35\times 10^{4}$ times larger than the \ion{Si}{2}] (2335\AA) line.
Since the equivalent widths of this \ion{Si}{1} line are at most five times
that of the \ion{Si}{2}] line toward any star, we see that neutral silicon
is $<$0.1\%\ of the total silicon in these sight lines.

The excited states of \ion{Fe}{2} and \ion{Si}{2} can be populated
by either collisions (with \ion{H}{1} and/or electrons) and/or by photon
pumping by ultraviolet photons
\citep{bw1968}. For only two of the six sight lines
studied do we see significant absorption due to the excited state
transitions of either \ion{Si}{2} or \ion{Fe}{2}: toward the stars HD~37021
and HD~37061.
Since there are lines arising in multiple excited states of \ion{Fe}{2}
in our STIS E230H datasets we first measured the column densities or upper
limits for states of \ion{Fe}{2} with excitation potentials up to $3117.5$
cm$^{-1}$ (0.39 eV) toward both HD~37021 and HD~37061.  We note that
toward HD~37061 we detect absorption for states up to
the $1872.6$ cm$^{-1}$ (0.24 eV) level, while toward HD~37021 only the
$1872.6$ cm$^{-1}$ level has detectable absorption.
Such lines have been observed previously in a range of contexts: e.g.,
circumstellar disks \citep{lagrange1996}, 
ejecta from $\eta$  Carina \citep{gull2005}, and circumstellar
outflow from the progenitor of SN 1998S \citep{bowen2000}.
We derived total column densities for the sum of all of these states of
$\rm\sim 5\times 10^{12}$ cm$^{-2}$ and $\rm\sim 2\times 10^{13}$ cm$^{-2}$
for HD~37021 (summed over both components, see below) and HD~37061,
respectively, which imply that $<$1\% of the \ion{Fe}{2} column is in
these excited states toward both stars.

The situation for \ion{Si}{2} in these two sight lines is more complicated
-- the non-detection of the 2335.321 \AA\ line of \ion{Si}{2}$^{\star}$ (the
287.2 cm$^{-1}$ (0.036 eV) level) in our
E230H datasets does not place a significant constraint on the column density,
the limits are $\sim$64\% and $\sim$35\% of the \ion{Si}{2} column for
HD~37021 and HD~37061, respectively.   The lines in the E140H datasets
for these stars are also not useful as all five lines of \ion{Si}{2}$^{\star}$
between 1194 and 1309 \AA\ are highly saturated.  Since we cannot directly
measure the column density of \ion{Si}{2}$^{\star}$, we turn to
models using the excitation of \ion{Fe}{2} and \ion{C}{2} to estimate the
contribution of \ion{Si}{2}$^{\star}$ to N(\ion{Si}{2}).

First we considered the HD~37061 sight line where we can use the columns of
the \ion{Fe}{2} 3d$^6$($^5$D)4s a$^6$D levels as well as the columns of
\ion{C}{2} and \ion{C}{2}$^{\star}$ from \citet{sofia2004} to derive
an electron density which can be used to estimate the \ion{Si}{2}$^{\star}$
in this absorption component.  For \ion{Fe}{2} we used the two lowest energy
states (0.385 and 0.667 eV) of the 3d$^6$($^5$D)4s a$^6$D level, where we
detected four lines for each state, and derived column densities of
1.7$\pm 0.1\times 10^{12}$ cm$^{-2}$ and 0.6$\pm 0.2\times 10^{12}$ cm$^{-2}$ respectively.
For \ion{C}{2} we used the ratio of the excited to the ground--state
(N(\ion{C}{2}$^{\star}$)/(N(\ion{C}{2})$\sim$0.5) measured by
\citet{sofia2004}.  Using the measured column density ratios and the models
of the excitation of \ion{C}{2} and \ion{Si}{2} from \citet{sp1979} 
and of \ion{Fe}{2} from \citet{keenan1988}, we see that for
temperatures of $\sim$1--3,000 $K$ (consistent with the inferred b-value)
N(\ion{Si}{2}$^{\star}$)/N(\ion{Si}{2}) $\la$ 10\% for this sight line.
At these temperatures the excitation of \ion{C}{2} suggests that
${\rm n_H\sim 10^{3}}$ cm$^{-3}$ and ${\rm n_e\sim 10}$ cm$^{-3}$, implying
that hydrogen is $\sim$1\%\ ionized in this component.

The excitation equilibrium arguments given above for HD~37061 suggest 
the \ion{Si}{2} and \ion{Fe}{2} absorption detected in these weak lines
corresponds to the bulk of the \ion{H}{1} along this sight line, and does
not include a large component associated with an \ion{H}{2} region(s).  
A similar argument holds for the other sight lines, and further evidence
of the lack of a large \ion{H}{2} region contribution to these columns
is the generally good agreement between the observed \ion{C}{2},
\ion{O}{1}, and \ion{Kr}{1} columns in these sight lines and the local
interstellar mean values.  Except for HD~152590, the sight lines
show no evidence for significant enhancements of the \ion{C}{2} column
suggestive of potential \ion{H}{2} region contamination \citep{sofia2004}.
For HD~152590 the column density of \ion{C}{2} (and perhaps \ion{Kr}{1})
is significantly larger than the mean value, this sight line is the only
sight line in our sample at large distances \citep{cartledge2001,sofia2004}.

\subsection{Column Densities}

We measured the equivalent widths or upper limits of the \ion{Fe}{2}
$\lambda\lambda$ 2234, 2249, 2260, 2344, 2367, 2374, 2382 \AA\ transitions
in our {\it HST} data, as well as the \ion{Si}{2}] $\lambda$ 2335 \AA\
intersystem transition (see Figure 1).  For four of these stars (all but
HD 37021 and 37061) we used {\it FUSE} data to measure the equivalent widths
of the \ion{Fe}{2} $\lambda\lambda$ 1055, 1142, 1143, and 1144 \AA\
lines.  With the exception of the 2234 and 2367 \AA\ transitions,
we adopted  the oscillator strengths from \citet{morton2003} to derive
curves-of-growth for each sight line. The equivalent width and the corresponding
optically thin column density for each of the above lines toward
HD~147888 are presented in Table~2, with the resulting curve--of--growth
for HD~147888 shown in Figure~2. From this plot, it is clear that
$\lambda\lambda$1144, 2344, 2374, 2382 are all extremely optically thick
lines, and that the \citet{ru1998} $f$-value for 2234 \AA\ 
seems to underestimate the \ion{Fe}{2} column density. 

For the \ion{Fe}{2} $\lambda$2367 \AA\ line there are several different
$f$-values in the literature.  The value given in \citet{morton2003} 
(log $f\lambda$ = -1.291) is based upon theoretical calculations, while
that given by \citet{cs1995} (log $f\lambda$ = -0.827) was
empirically derived from observations of several interstellar sight lines.
\citet{welty1999} suggested revising the \citet{cs1995} oscillator strength
upwards to reflect revisions in the oscillator strengths of the other
\ion{Fe}{2} lines used by \citet{cs1995} to determine that value.  We adopt
the $f$-value from \citet{welty1999} (log $f\lambda$ = -0.713) for 
$\lambda$2367 \AA\ based on the curves-of-growth for the sight lines in
this study.  Were we to adopt
the \citet{morton2003} $f$-value the measured equivalent widths of this line
would imply that the \ion{Fe}{2} columns toward HD 37021, HD37061, and HD 147888
should increase by factors of 0.53, 0.54, and 0.53 dex, respectively.
Similarly, use of the \citet{cs1995} oscillator strength would imply
an increase the Fe column densities for all three sight lines by 0.11 dex.
For the stars HD 152590 and HD 207198 it appears that the profiles of
the 2367 \AA\ transition fail to fully reflect the multiple distinguished
absorption components in these sight lines \citep{cartledge2001} based upon
the \ion{Fe}{2} 2249 \AA\ transition.  We note that the column densities,
based on direct measurements of the \ion{Fe}{2} 2249 \AA\ transition
for these sight lines may be underreported due to saturation.  For HD 27778
we do not detect the \ion{Fe}{2} 2367 \AA\ transition at a statistically
significant level, and therefore instead adopt the value derived from the
apparent optical depth integral for the 2249 and 2260 \AA\ transitions which is
consistent with the column derived from the curve-of-growth.
In addition, we have included the error due to uncertainties in the oscillator
strengths in deriving the uncertainties in the resulting \ion{Fe}{2} column
densities based upon the curve--of--growth.  Errors were taken from
\citet{morton2003} for the STIS lines except 2367 \AA\ where we
have used the error estimate of \citet{cs1995}, for the {\it FUSE} lines
we have used the error estimates from \citet{howk2000}.
We use the equivalent width of the \ion{Si}{2}] 2335 \AA\ line to directly
derive the column density for \ion{Si}{2} from the weak line limit. The
uncertainty in the equivalent widths take into account both photon
noise \citep{jenkins1973} and the uncertainty in the continuum placement
\citep{ssm1991, scs1992} added in quadrature.  In Table 1                   
we list a representative \ion{Fe}{2} transition used, the equivalent widths
($W_\lambda$) for the listed \ion{Fe}{2} line and \ion{Si}{2}] 2335\AA,
and the respective column densities ($N$) for our sight lines and $\zeta$ Oph.
We have also separately listed the errors in the \ion{Si}{2} column density
including the uncertainty in the oscillator strength from \citet{morton2003}
added in quadrature -- while not relevant for intercomparisons between
sightlines using the same (weak) transition, such errors can be relevant for
intercomparisons between sightlines using different absorption lines.

\subsection{\ion{Fe}{2} $\lambda$2234 \AA\ Detection}

The ($^4$F$^0_{10}$) \ion{Fe}{2} $\lambda$2234 \AA\ transition was detected
toward three sight lines: HD 37021, HD 37061, and HD 147888. Each of these
detections was at the 4 $\sigma$ level or better. Figure 3 shows the normalized
flux of the 2234 \AA\ ({\it solid line}) absorption feature toward each of these
targets. The normalized flux of the 2367 \AA\ ({\it dotted line}) absorption
feature has been superimposed over these data with a different vertical
scale. The figure suggests that, while the 2234 \AA\ feature is very weak,
it is present and detectable along high column density sight lines.

The only available $f$-value for 2234 \AA\ (log $f\lambda$ = -1.249;
\citealt{morton2003}) is based on the theoretical work of \citet{ru1998}.
Were we to adopt this oscillator strength the
Fe {\sc ii} column densities for HD 37021, HD 37061, and HD 147888 would
decrease by 0.15 dex, 0.37 dex, and 0.34 dex, respectively.  By fitting
the measured equivalent widths for 2234 \AA\ to the curves of growth for
HD 37021, HD 37061, and HD 147888, we have instead derived an empirical
$f$-value of $log f\lambda$ = -1.54 $\pm 0.05$ for this transition. 
For each sight line we first determined the column density and associated
error using a curve--of--growth, we then calculated what the corresponding
oscillator strength (as well as its error) would be given the observed
equivalent width and error for the 2234 \AA\ line.  The derived oscillator
strength was then determined by taking the weighted mean for each sight line 
based upon equivalent width measurement uncertainties.  Our estimate
of the uncertainty is the error in the weighted mean from these
three determinations.

\section{Discussion}

Although dust grains containing silicon and iron are common in the ISM, it
is not clear what their size distributions or origins are
\citep{zubko2004}. Using a simplistic dust model, \citet{sofia2005} find
that sight lines with larger mass-fractions in small grains have a higher
percentage of their silicon in the form of dust, and that the "extra"
silicon is depleting primarily into small particles with effective
radii $< 200\ \micron$. Although the model suggests
that an additional population rather
than grain processing is responsible for the enhanced small-grain
populations, the authors warn that their result must be interpreted with
caution. The model that they used was quite simple, so their results were
meant to be merely suggestive.

The present data set represents the first sample of high-quality
interstellar gas-phase Si and Fe measurements in sight lines exhibiting
substantial levels of extinction and having measured $c_{4}$ and $R_{V}$
values \citep{valencic2004}. Although our data only sample a small number
of targets, these sight lines and measured abundances are ideal for
exploring the small grain population.  However, we must first convert 
gas--phase abundances to dust-phase abundances and/or depletions.
Typically one adopts as the reference for the current total (dust $+$ gas)
abundances the Solar--system values, since the required set of cosmic
abundance standards for current interstellar matter in and around the
Solar--neighborhood is not currently established \citep{ss1996}.  We note
that it does appear that the interstellar
medium is well--mixed, since the variations between sight lines in the
interstellar gas--phase abundances of elements such as \ion{O}{1} and
\ion{Kr}{1} are small (see for example \citet{cartledge2003, cartledge2004}).
Thus if the ``true'' interstellar reference abundances are different than the
proto--sun reference abundances of \citet{lodders2003} used here,
the differences will amount to just a uniform offset in the measured number
of atoms in the dust--phase for silicon and iron for these sight lines.

One measure of the small grain population is the $c_{4}$ variable in the
\citet{fm1988} extinction parameterization, which
describes the strength of the extinction from 1050 to 1695 \AA\
\citep{sofia2005}. Specifically, \citet{fm1988} found a polynomial common
to extinction curves that fits $E(\lambda-V)/E(B-V)$ as a function of
wavenumber below 1695 \AA. The $c_{4}$ parameter is the amplitude of this
extinction component for a given sight line. The polynomial is a
continuously decreasing function from 1050 to 1695 \AA, therefore larger
$c_{4}$ values correspond to steeper extinction curves over these
wavelengths. Since extinction in this wavelength region is produced by
small grains either composed of carbon \citep{mrn1977,desert1990} or
silicates \citep{mathis1996,ld2001,clayton2003}, $c_4$ can be used as a
proxy for grain size distribution; larger $c_{4}$ values correspond to a
higher fraction of small grains along a sight line.
The ratio of total-to-selective extinction, $R_V = A_V/E(B-V)$,
can also serve as a proxy for dust size distributions with smaller
$R_V$ values indicating a larger fraction of mass in small grains. In
Figure 4 we display the silicon and iron dust-phase abundances as a
function of sight line $c_{4}$ values. The figure shows that in
higher-$c_{4}$ sight lines a smaller fraction of Si and Fe atoms are in the
gas phase, suggesting that sight lines with enhanced populations of small
grains have more of these elements incorporated into dust. Using $R_{V}$ as
a proxy for dust-size distributions produces a similar result, however
there is no correlation between depletions and {\it E(B-V)} in our sample.
This suggests
that the dust-phase abundance variations of Si and Fe are truly related to
grain-size populations rather than to the strength of extinction. A linear
fit to the Si data in Figure 4 (a higher order polynomial is not
statistically justifiable) indicates that there is a 6\% rise in the
silicon dust-phase abundance per increase of 1.0 in $c_{4}$.
\citet{sofia2005} found that in order to fit the extinction in their
sample, their model required 12\% more silicon to be in dust per 1.0
increase in $c_{4}$. The quantitative difference in these results is likely
due to the simplicity of the model used to fit the extinction.

There are two plausible, non-exclusive explanations for the correlation
between depletions and $c_{4}$ (or $R_{V}$). The first is that the excess
of elements depleted from the high-$c_{4}$ sight lines is incorporated into
a new population of small grains; these grains would exist in addition to
the grains found in lower-$c_{4}$ regions. Alternatively, a grain
population that has been processed to a smaller average size could result
in enhanced depletions. The increased surface-to-volume ratio would augment
the number of surface sites available, and thus allow enhanced mantling.

Carbon appears to be under-abundant in the gas toward HD 27778
\citep{sofia2004},
which also shows the most extreme Si and Fe depletions. The least-depleted
sight lines in our sample, toward HD 37021 and HD 37061 (both in Orion),
show low molecular fractions and significantly higher radiation fields. The
physical characteristics of these sight lines suggest that the measured
depletions may be due to differences in the survival of small grains in
some interstellar environments, perhaps coupled with additional mantling
onto grain surfaces. We note that \citet{lw2005} also find that enhanced
small grain populations (as indicated by low $R_{V}$ values) are common in
high--latitude translucent-cloud sight lines where one might expect
relatively low radiation fields.

Figure 5 shows dust-phase silicon abundances plotted as a function of
dust-phase Fe abundances for regions that sample a wide variety of neutral
ISM environments, from halo clouds to the translucent sight lines of our
sample. The solid line in the figure shows the weighted linear least
squares fit to the data; the slope is 3.46 and the y-intercept is -80.42.
Statistically this is a valid fit to the points, with 69\% of the 1$\sigma$
error bars intersecting the line. We note, however, that those data points
with dust-phase Fe values below 34 per million H (PMH) could be better fit
with a less-steep line (shown as a dashed line in Figure 5); a fit 
to the points with (Fe/H)$_{dust}$ greater
than 34 PMH produces a line that is similar to the solid one plotted in
the figure. Given the small number of points in Figure 5 and the large
uncertainties of the data at smaller dust abundances, we cannot conclude
with certainty whether a one or a two line fit is appropriate for the
plotted data. 
We note that in neither case could the illustrated linear fits extend to
indefinitely low values of dust abundances; both would have more than half
of the iron in dust when silicon is entirely in the gas phase.
At some point Fe has to deplete more efficiently from the gas than Si, which would
result in a less steep relationship than that shown by the lines in Figure
4. 

The one and two line fits to the Figure 5 data result in different
interpretations for the small grain population.
The single-line scenario suggests that over the large range of physical
conditions sampled by the sight lines, Si and Fe are being
depleted into dust in a ratio that is equal to the line's slope, 3.46 Si
atoms to every Fe atom. Further, if the differences in elemental
dust--phase abundances among
neutral interstellar sight lines is the result of mantling
\citep{scs1994,barlow1978} then this linear trend coupled with the
correlations in Figure 4 implies that the chemical composition of the small
grain population closely resembles that of mantles.
Conversely, if the two-line fit is appropriate for the data in Figure 5,
this would imply that the small-grain population incorporates a higher
Si-to-Fe ratio than mantles, suggesting that the small grains may be silicon
rich, which complements the model results of \citet{sofia2005}. A larger
number of well-measured interstellar Si and Fe abundances are needed to
distinguish between these two interpretations for the small grains.

\acknowledgements
We gratefully acknowledge partial support from Illinois NASA Space 
Grant Consortium grant number NGT5-40073 subcontract to Northwestern
University, and the Space Telescope Science Institute grant GO-9465.01 to
Whitman College.

The authors would like to thank the referee, Dr. Jeffrey Linsky, for his
helpful comments that have improved this paper.  We would also like to  thank
David Knauth for his discussions on the material of this paper.

\clearpage

\begin{deluxetable}{lcccccccccccc}
\setlength{\tabcolsep}{0.025in}
\tabletypesize{\scriptsize}
\tablecolumns{13}
\tablecaption{Sight Line Characteristics with Iron and Silicon Abundances\label{tbl-1}}
\tablewidth{0pt}
\tablehead{
 & & & & & \multicolumn{4}{c}{Fe {\tiny II}} & & \multicolumn{3}{c}{Si {\tiny II}\tablenotemark{c}} \\
\cline{6-9} \cline{11-13} \\
\colhead{} & \colhead{$E(B-V)$\tablenotemark{a}} & \colhead{} & \colhead{} & \colhead{$N$(H)\tablenotemark{a}} & \colhead{Line Listed} & \colhead{$W_\lambda$} & \colhead{$N$} & & & \colhead{$W_\lambda$} & \colhead{$N$} & \\
\colhead{Star} & \colhead{(mag)} & \colhead{$R_V$\tablenotemark{b}} & $c_4$\tablenotemark{b} & \colhead{(10$^{21}$ cm$^{-2}$)} & \colhead{(\AA)} & \colhead{(m\AA)} & \colhead{($10^{14}$ cm$^{-2}$)} & $\delta$(Fe)\tablenotemark{c} & & \colhead{(m\AA)} & \colhead{($10^{15}$ cm$^{-2}$)}\tablenotemark{d} & $\delta$(Si)\tablenotemark{c}
}
\startdata
HD 27778 & 0.36 & 2.6 & 1.00 & 2.3$\pm$0.4 & 2260 & 24.7$\pm$0.5 & 2.61$\pm$0.05 & -2.49$\pm$0.18 & & $<$2.45\tablenotemark{e} & $<$8.62\tablenotemark{e} & $<$-1.04\\
HD 37021 & 0.54 & 5.8 & 0.04 & 4.8$\pm$1.1 & 2367 & 10.4$\pm$0.6 & 27.9$\pm$1.5 & -1.78$\pm$0.24 & & 2.83$\pm$0.64 & 14.3$\pm$3.2 & -1.14$\pm$0.14\\
HD 37061 & 0.45 & 4.3 & 0.21 & 5.4$\pm$1.1 & 2367 & 10.3$\pm$0.4 & 27.6$\pm$1.0 & -1.83$\pm$0.21 & & 2.89$\pm$0.36 & 14.4$\pm$1.8 & -1.18$\pm$0.11\\
HD 147888 & 0.52 & 3.9 & 0.34 & 5.9$\pm$0.9 & 2367 & 8.39$\pm$0.52 & 23.1$\pm$1.4 & -1.95$\pm$0.16 & & 2.58$\pm$0.47 & 13.1$\pm$2.4 & -1.26$\pm$0.37\\
HD 152590 & 0.39 & ... & ... & 2.9$\pm$0.3 & 2249 & 93.4$\pm$2.4 & 15.5$\pm$0.4 & -1.81$\pm$0.11 & & 1.68$\pm$0.55 & 8.31$\pm$2.71 & -1.15$\pm$0.0.48\\
HD 207198 & 0.59 & 2.8 & 0.77 & 4.8$\pm$1.1 & 2249 & 78.8$\pm$1.2 & 13.5$\pm$2.1 & -2.09$\pm$0.28 & & 1.40$\pm$0.68 & 6.91$\pm$3.35 & -1.45$\pm$0.26\\
$\zeta$ Oph\tablenotemark{f} & 0.60 & 2.6 & 0.56 & 1.4$\pm$0.1 & 2249 & 22.5$\pm$1.1 & 3.09$\pm$.14 & -2.20$\pm$0.09 & & 0.48$\pm$0.12 & 2.89$\pm$0.68 & -1.30$\pm$0.34\\
\enddata
\tablenotetext{a}{\citet{cartledge2004}.}
\tablenotetext{b}{\citet{valencic2004}.}
\tablenotetext{c}{Inferred logarithmic depletions of Fe and Si into the
dust--phase based upon the derived column densities, the total hydrogen column
densities listed, and the the proto--solar abundance value for Fe and Si
from \citet{lodders2003}.  \ion{Si}{2} columns derived from the 2335\AA line.}
\tablenotetext{d}{Errors listed include continuum placement and photon noise
errors only, errors including the uncertainty in oscillator strength are
(in units of $10^{15}$ cm$^{-2}$) 4.1, 3.2,3.4, 3.10, 3.58, and 0.89
for the six detections respectively.}
\tablenotetext{e}{3 $\sigma$ limit.}
\tablenotetext{f}{The Fe {\tiny II} $W_\lambda$ is from \citet{scs1992}. \citet{scs1992} used now outdated $f$-values in their determination of the Fe {\tiny II} column density towards $\zeta$ Oph. This value has been changed to reflect the current $f$-value for the 2249 \AA\ transition \citep{howk2000}. The Si {\tiny II} $W_\lambda$ is from \citet{cardelli1994}. $\zeta$ Oph has two absorbing components, but \citet{cardelli1994} could only measure the 2335 \AA\ transition for the stronger component B. \citet{scs1994} report the column density for component A, based on measurements of Si {\tiny II} $\lambda$1808 \citep{scs1992}, and component B. The two components have been combined here.}
\end{deluxetable}

\clearpage

\begin{deluxetable}{cccc}
\tabletypesize{\scriptsize}
\tablecolumns{4}
\tablecaption{EQWs and Optically--thin Column Densities for HD 147888\label{tbl-2}}
\tablewidth{0pt}
\tablehead{
\colhead{Fe~II line} & \colhead{$log (\lambda$$f)$} & \colhead{$W_\lambda$} & \colhead{$N$} \\
\colhead{(\AA)} & & \colhead{(m\AA)} & \colhead{(cm$^{-2}$)}}
\startdata
1055  &   0.81  &    28.2$\pm$3.8 & (5.43$\pm$0.74) $\times 10^{14}$ \\
1142  &   0.66  &    42.2$\pm$9.9 & (1.06$\pm$0.25) $\times 10^{15}$ \\
1143  &    1.34 &    38.9$\pm$5.3 & (2.22$\pm$0.30) $\times 10^{14}$ \\
1144  &    1.98 &     105$\pm$5   & (1.40$\pm$0.06) $\times 10^{14}$ \\
2234  &   -1.54 &    1.20$\pm$0.3 & (1.10$\pm$0.25) $\times 10^{15}$ \\
2249  &   0.61  &    44.2$\pm$0.8 &  (1.06$\pm$0.02) $\times 10^{15}$ \\
2260  &   0.74  &    59.3$\pm$0.8 &  (9.87$\pm$0.13) $\times 10^{14}$ \\
2367  &   -0.71 &   8.39$\pm$0.52 &  2.31$\pm$0.14 $\times 10^{15}$ \\
2344  &    2.43 &     227$\pm$1   & (1.41$\pm$0.004) $\times 10^{14}$ \\
2374  &    1.87 &     163$\pm$1   & (2.80$\pm$0.01) $\times 10^{14}$ \\
2382  &    2.88 &     266$\pm$1   & (7.38$\pm$0.03) $\times 10^{13}$ \\
\enddata
\end{deluxetable}

\clearpage

\begin{figure}
\plotone{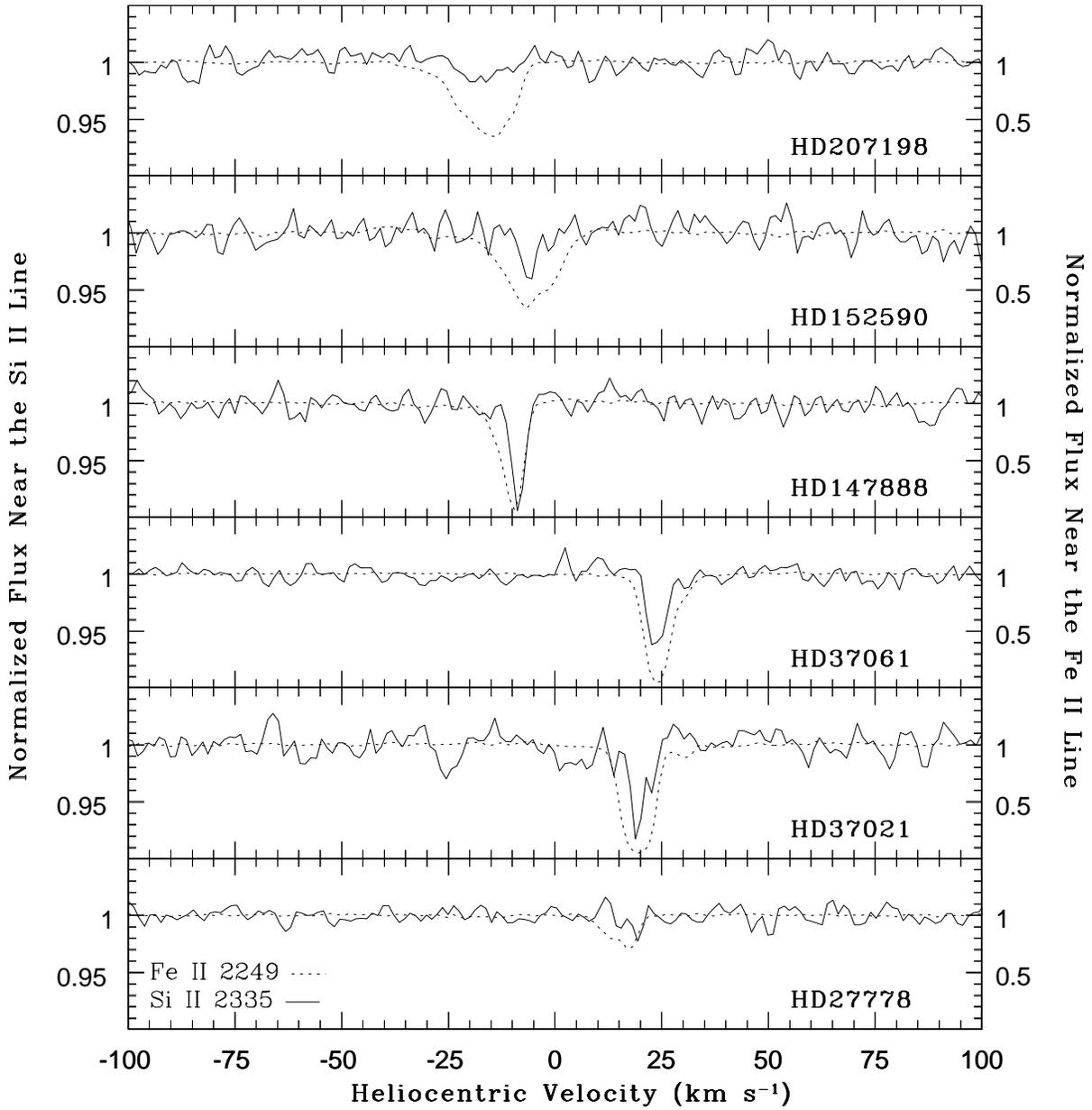}
\caption{Normalized STIS echelle spectra of the \ion{Si}{2}] $\lambda$2335
({\it solid line}) and \ion{Fe}{2} $\lambda$2249 ({\it dotted line})
absorption features. Note that the normalized flux scale is on the left for
silicon and on the right for iron.\label{fig-1}}
\end{figure}

\clearpage

\begin{figure}
\epsscale{.80}
\plotone{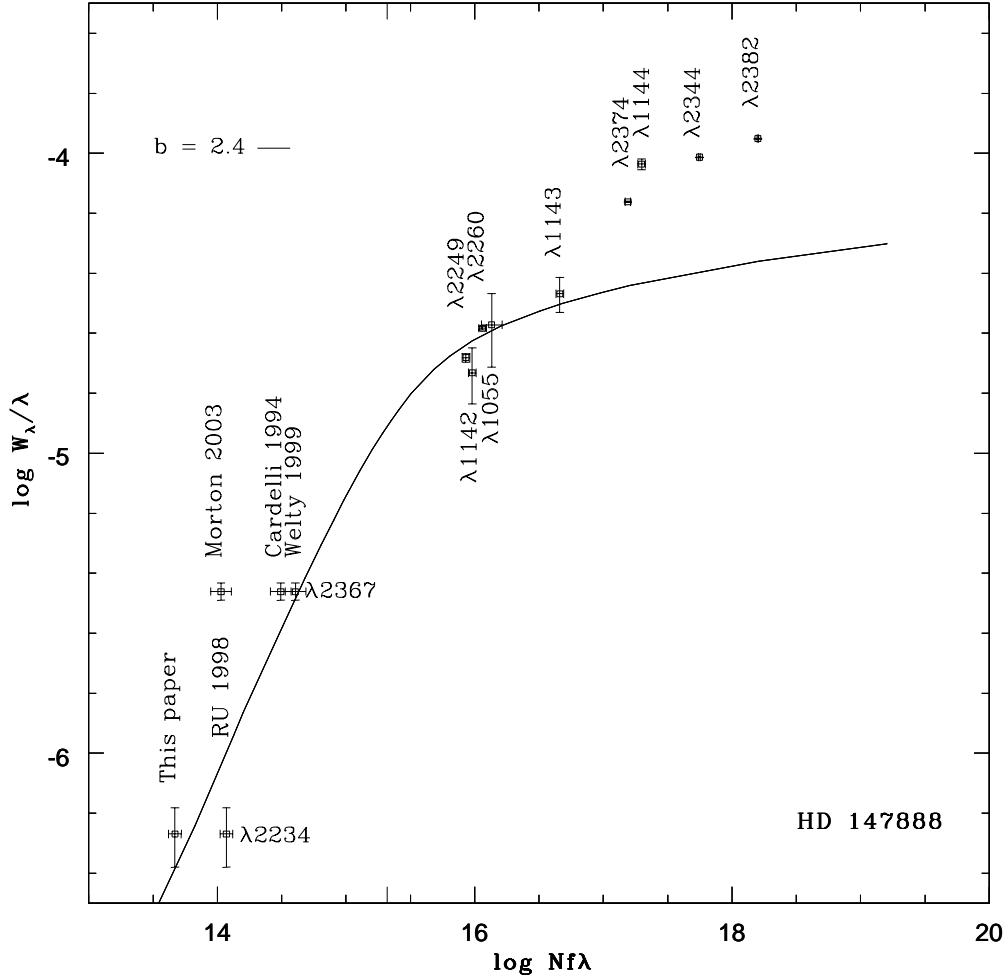}
\caption{Curve of growth for \ion{Fe}{2} absorption lines toward HD 147888,
based on {\it STIS} and {\it FUSE} data.  The lines at $\lambda$\ 1144,
2344, 2374, 2382 \AA\ have appear to have larger equivalent widths than would
be expected based on the low $f$--value lines, because there are a few low
column density components along this sight line that are not measurably
represented in the weaker lines. These low column density components are
outside the main component(s) seen in Figure 1 (with a best fit doppler
width $b=2.4$ km s$^{-1}$), but do provide
a significant addition to the equivalent widths of the stronger lines, and
 therefore these lines fail to follow the curve of growth.  These components are
however, justifiably ignored for the purposes of this analysis due to their
low total column and lack of corresponding \ion{Si}{2}] 2335 \AA\ absorption.
The three different oscillator strengths for  \ion{Fe}{2} 2367 \AA\ are shown,
showing that the \citet{welty1999} value is the best choice.  Both oscillator
strengths for  \ion{Fe}{2} 2234\AA, the one calculated by \citet{ru1998} and
the one determined in this paper, are shown.  The $f$--value calculated by
\citet{ru1998} appears to underestimate the total \ion{Fe}{2} column density.
Uncertainties in the oscillator strengths are shown by the horizontal error
bars (see text).
}
\end{figure} 

\clearpage

\begin{figure}
\epsscale{1.0}
\plotone{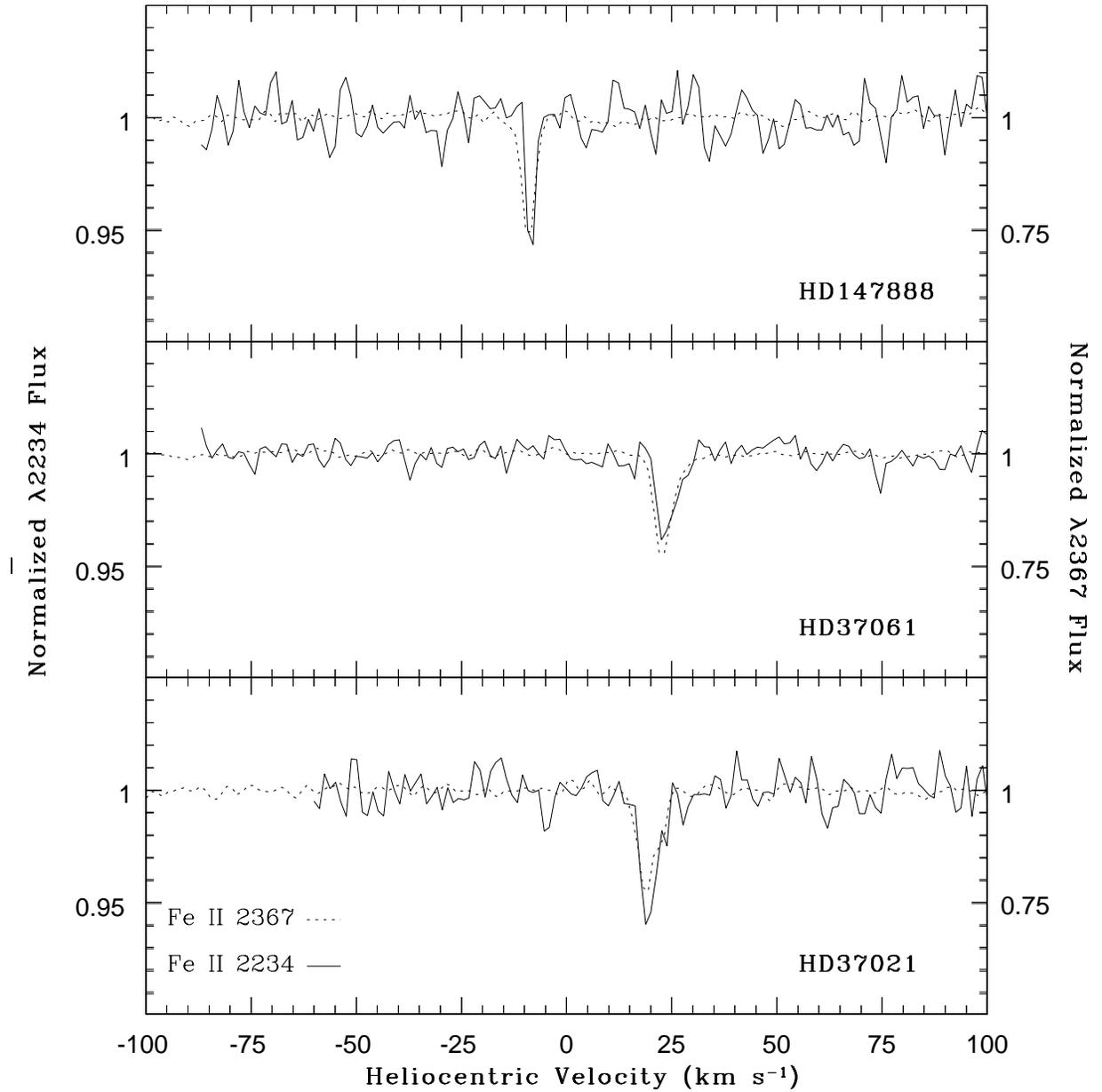}
\caption{Normalized STIS echelle spectra of the \ion{Fe}{2} $\lambda$2234
({\it solid line}) and \ion{Fe}{2} $\lambda$2367 ({\it dotted line})
absorption features. Note that the normalized flux scale is on the left for
the 2234 \AA\ and on the right for the 2367 \AA\ data. \label{fig-2}}
\end{figure}

\clearpage

\begin{figure}
\plotone{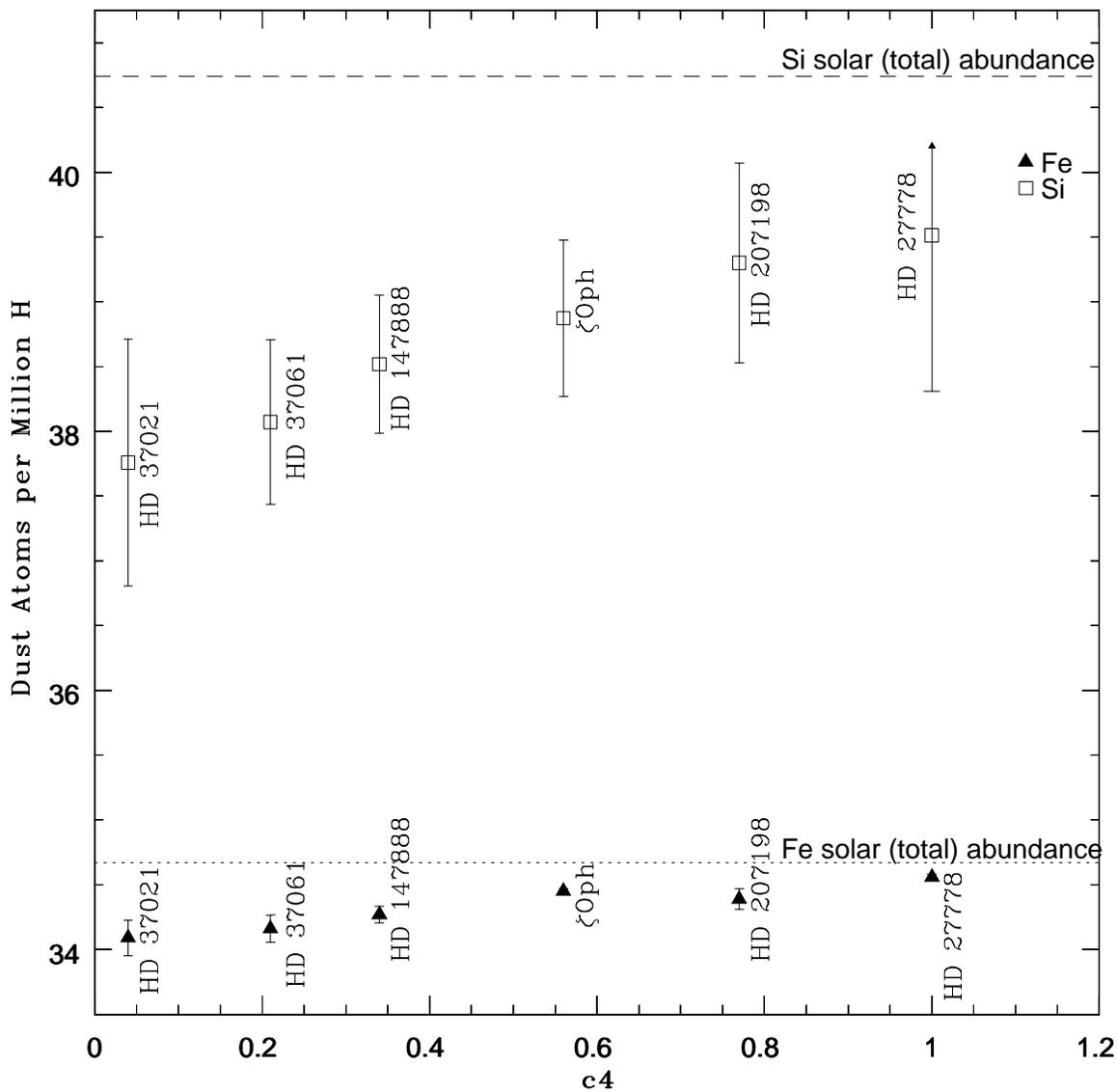}
\caption{Number of silicon and iron atoms in the dust per $10^6$ H versus
the \citet{fm1988} extinction coefficient $c_4$, with 1 $\sigma$ error bars.
The $c_4$ parameter describes the strength of the rise toward the
far--ultraviolet, and is believed to be related to the abundance of small
grains. Note that the iron uncertainties for $\zeta$ Oph and HD 27778 are less
than the size of the point on the plot. The proto--solar abundance values
are from \citet{lodders2003}.\label{fig-3}}
\end{figure}

\clearpage

\begin{figure}
\plotone{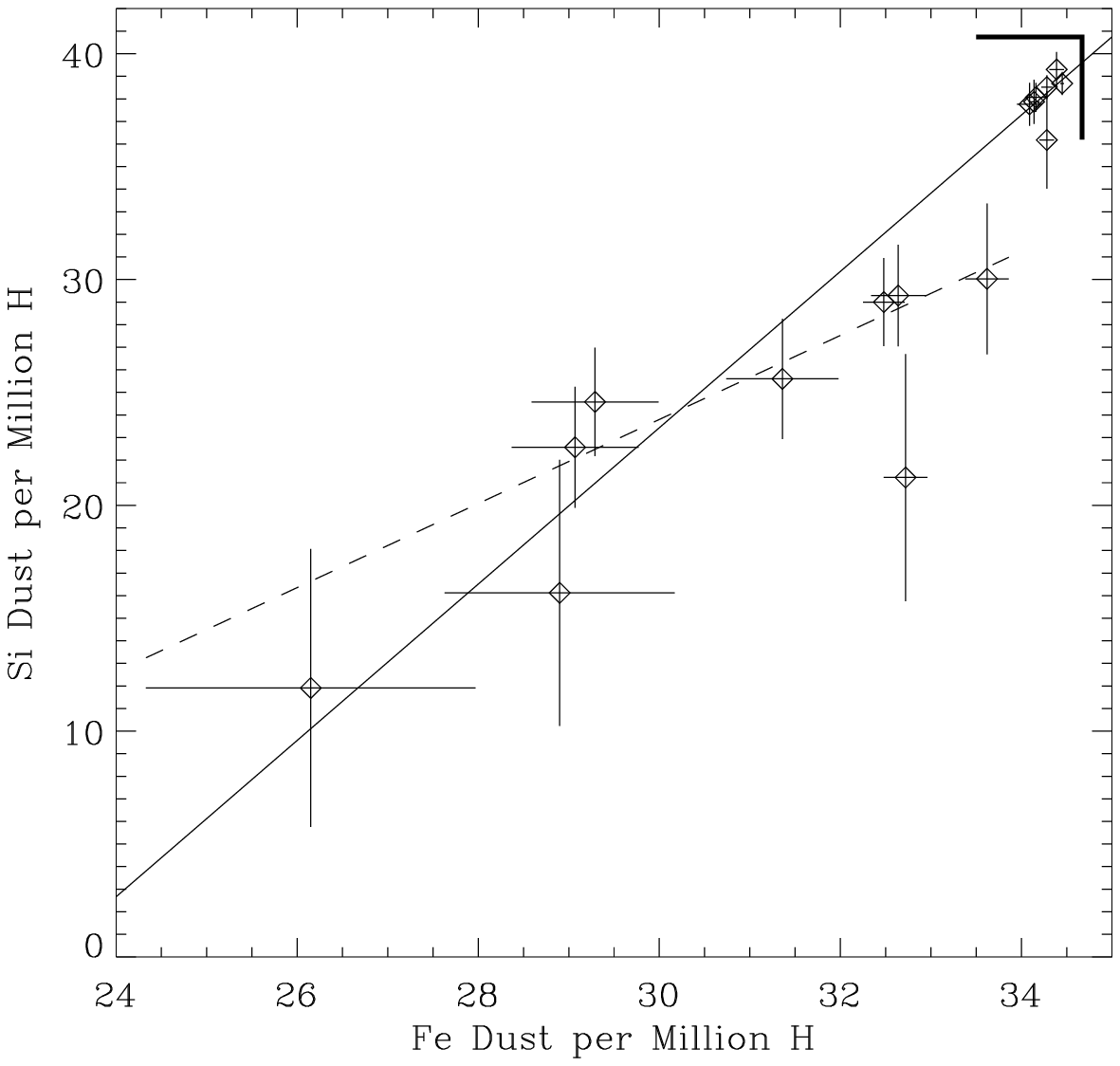}
\caption{Dust-phase Si nuclei per million H nuclei versus dust-phase Fe nuclei per million H nuclei, 
with 1 $\sigma$ error bars. The sight line 
(component, if necessary) - and source for each point on this plot from left to right 
are:  HD 149881(2) - \citet{sf1995}, HD 93521(3) - \citet{sf1993}, HD 93521(6), 
HD 149881(6), HD 18100 - \citet{ss1996}, HD 215733(19) 
- \citet{sf1997}, HD 93521(8), $\mu$ Col(1) - \citet{hsf1999}, 
23 Ori(WLV) - \citet{welty1999}, and the 7 points in the upper right hand 
portion of the plot are the sight lines from Table 1, minus HD 27778, and 
23 Ori(SLV) \citep{welty1999} which lies below the others. Note
that HD 27778 is not included 
in this plot as only an upper limit is available for Si. The solar abundance values 
for Si and Fe \citep{lodders2003} are shown by the bold lines in the upper right 
hand portion of the plot. The solid line is a least squares fit to all the data shown, 
while the dashed line represents a fit to only those points with less than 34 
iron atoms in the dust per million H atoms.}
\end{figure}

\end{document}